\documentclass{article}
\usepackage{spconf,amsmath,graphicx,hyperref,enumitem, algorithm, algorithmic}
\usepackage{amssymb}   

\usepackage[table]{xcolor}
\usepackage{adjustbox}

\definecolor{lightblue}{rgb}{0.88, 0.93, 0.97}
\definecolor{lightgray}{rgb}{0.95, 0.95, 0.95}

\title{SightSound-R1: Cross-Modal Reasoning Distillation from Vision to Audio Language Models}
\name{Qiaolin Wang$^\sharp$, Xilin Jiang$^\sharp$, Linyang He$^\sharp$, Junkai Wu$^\flat$, Nima Mesgarani$^\sharp$}
\address{$^\sharp$Columbia University, New York, NY, USA\\
$^\flat$University of Washington, Seattle, WA, USA}
\begin{document}
\ninept
\maketitle
\begin{abstract}

While large audio-language models (LALMs) have demonstrated state-of-the-art audio understanding, their reasoning capability in complex soundscapes still falls behind large vision–language models (LVLMs). Compared to the visual domain, one bottleneck is the lack of large-scale chain-of-thought audio data to teach LALM stepwise reasoning. To circumvent this data and modality gap, we present \textbf{SightSound-R1}, a cross-modal distillation framework that transfers advanced reasoning from a stronger LVLM teacher to a weaker LALM student on the same audio–visual question answering (AVQA) dataset. SightSound-R1 consists of three core steps: (i) test-time scaling to generate audio-focused chains of thought (CoT) from an LVLM teacher, (ii) audio-grounded validation to filter hallucinations, and (iii) a distillation pipeline with supervised fine-tuning (SFT) followed by Group Relative Policy Optimization (GRPO) for the LALM student. Results show that SightSound-R1 improves LALM reasoning performance both in the in-domain AVQA test set as well as in unseen auditory scenes and questions, outperforming both pretrained and label-only distilled baselines. Thus, we conclude that vision reasoning can be effectively transferred to audio models and scaled with abundant audio-visual data.

\end{abstract}
\section{Introduction}
\label{sec:intro}

The rapid progress of multimodal large language models \cite{qwen25vl,videollama3,qwen25omni,qwen2audio,Gpt-4o} has produced systems that couple modality-specific encoders (e.g., Whisper \cite{whisper} for audio and Vision Transformers \cite{ViT} for images) with powerful large language model decoders. These foundation models already demonstrate strong performance in audio and video understanding \cite{mmmu,mmau}. While supervised next-token prediction may only learn shallow answer patterns, reinforcement learning (RL) \cite{deepseekmath} and chain-of-thought (CoT) prompting \cite{CoT} have emerged as an effective path to learn reasoning, with DeepSeek-R1 \cite{deepseek} achieving substantial gains from pure RL and inspiring subsequent “R1-style’’ training in vision domains \cite{vision-r1,Visual-rft,r1-aqa}. Comparable but delayed progress has recently begun to appear in the audio domain \cite{Audio-cot,Audio-reasoner}. 

To measure and compare audio and visual reasoning in a concrete, challenging setting that demands temporal, comparative, and causal inference, we focus on audio–visual question answering (AVQA) \cite{avqa,musicavqa,aura}. Despite shared scenes, prior work observes a persistent gap between large vision–language models (LVLMs) and large audio–language models (LALMs) \cite{bridging}: LVLMs—typically larger and trained on abundant image–text data—produce stronger multi‐step reasoning, whereas LALMs—smaller and trained on scarcer audio–text data—often struggle to generate coherent thinking traces in complex auditory scenes \cite{r1-aqa}. The scarcity of audio CoT further limits the application of RL in the audio domain. Attempts to synthesize audio CoT via pairing audio captions with text‐only LLMs \cite{Audio-Thinker,audsemthinker,thinksound} could help, but cross‐modal reasoning distillation, given a large gap between LVLM and LALMs, remains under-  and worth to be explored. This motivates a central question: \textbf{\textit{Can reasoning be transferred from LVLM to LALMs to bridge the modality gap in the same audio–visual scenes?}}


We study this question by analyzing Qwen2.5-VL \cite{qwen25vl}, Qwen2.5-Omni \cite{qwen25omni}, and Qwen2-Audio-7B-Instruct \cite{qwen2audio} on AVQA dataset \cite{avqa}. We first identify the performance gap between LVLMs and LALMs and applied a simple test-time distillation approach to alleviate it. We further introduce \textbf{SightSound-R1}, a cross-modal reasoning-distillation framework in which a powerful LVLM supervises a student LALM by providing filtered reasoning trajectories and predicted labels on AVQA dataset \cite{avqa}. SightSound-R1 surpasses zero-shot inference and label-only distillation baselines on MMAU-mini sound track \cite{mmau} and MusicAVQA dataset \cite{musicavqa}, yielding structured reasoning steps (Figure~\ref{fig:comparsion}). Our contributions are three-fold:

\begin{figure}
    \centering   \includegraphics[width=1.02\linewidth]{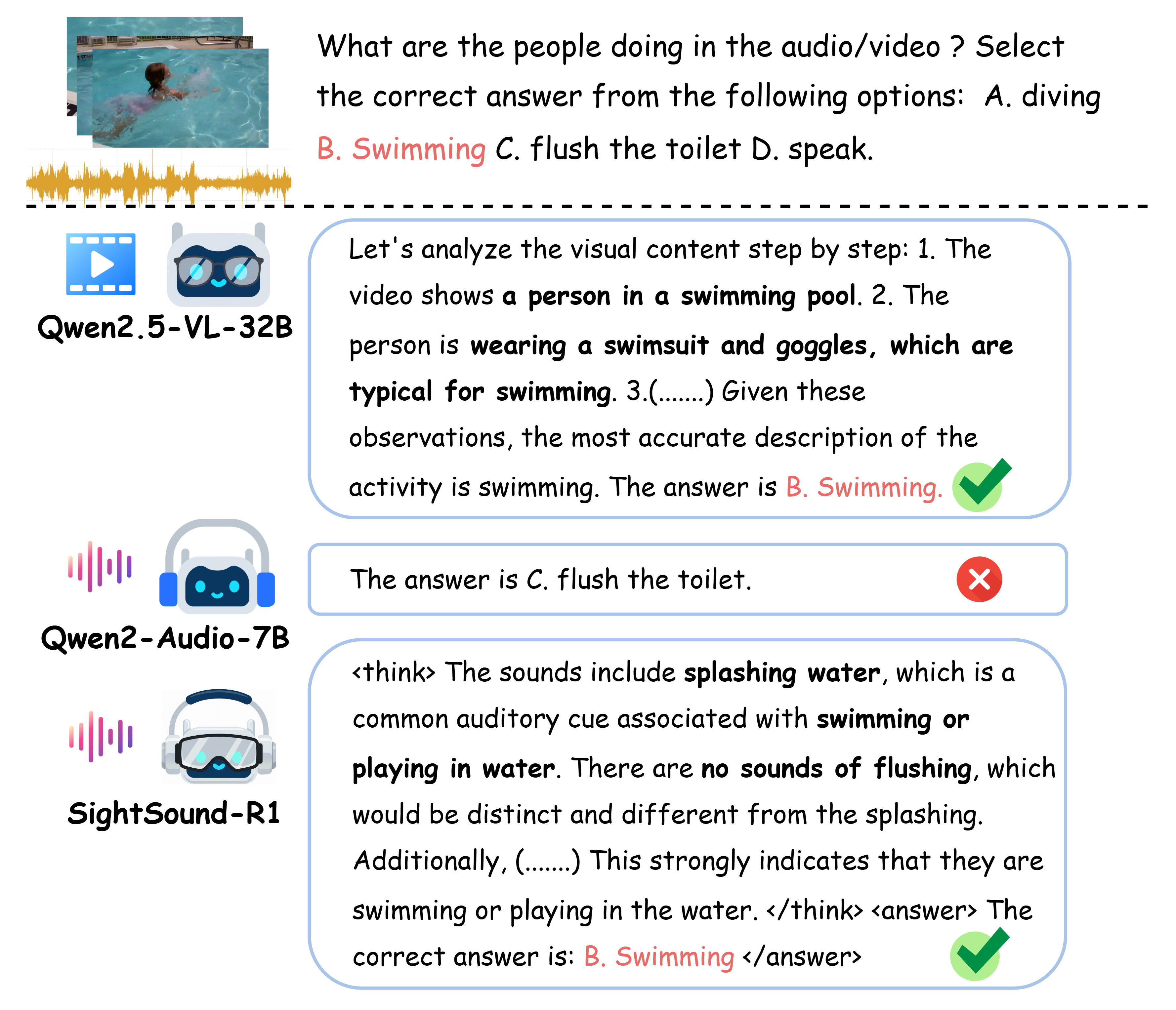}
    \caption{A strong LVLM (Qwen2.5-VL-32B) provides richer stepwise reasoning than the LALM (Qwen2-Audio-7B) on AVQA. SightSound-R1 further distills its audio-focused CoT traces into LALM to learn audio-grounded reasoning.}
    \label{fig:comparsion}
    \vspace{-0.5cm}
\end{figure}

\begin{figure*}[t]
    \centering
    \includegraphics[width=1\linewidth]{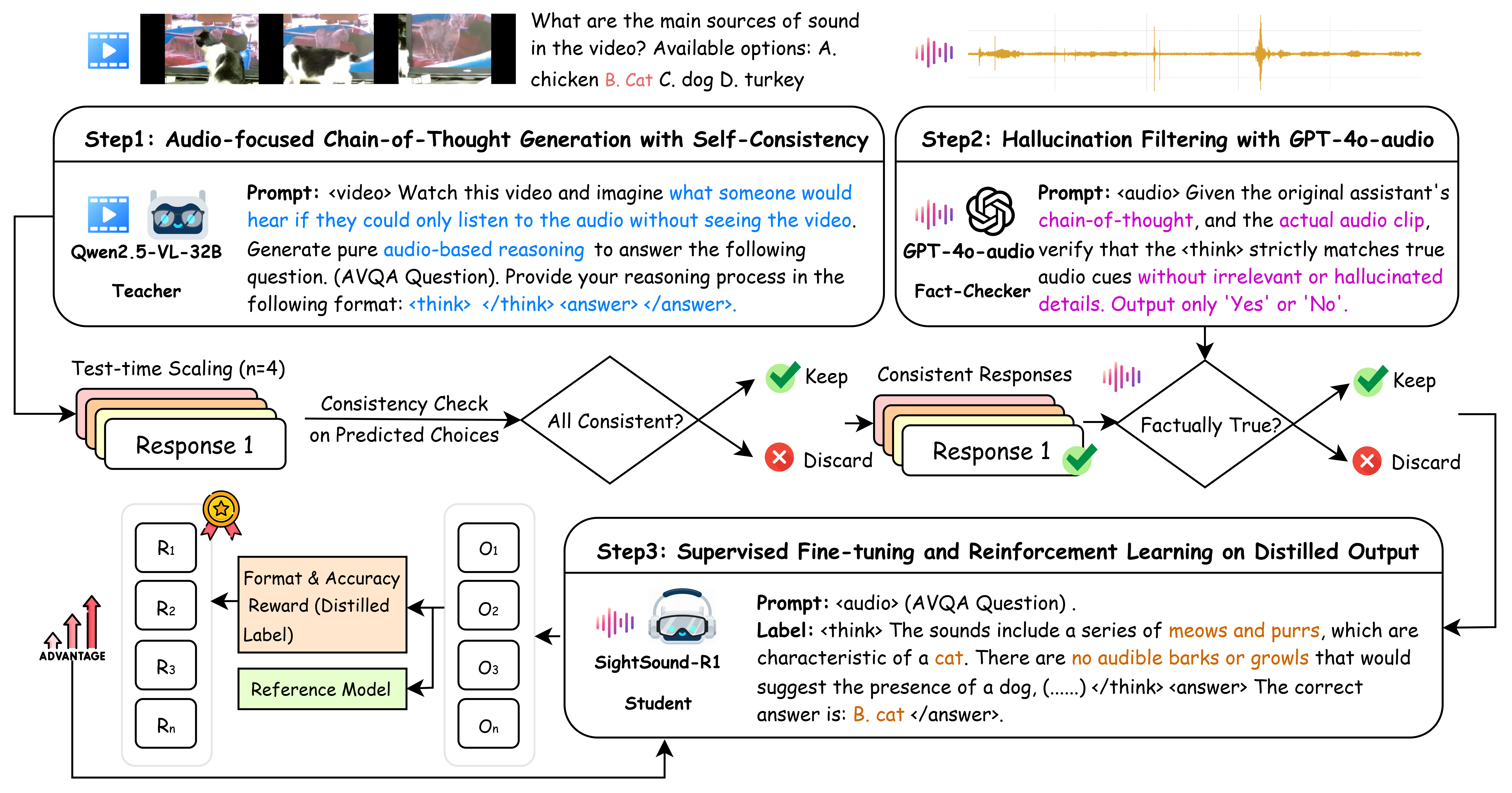}
    \caption{\textbf{Overview of SightSound-R1 framework}. The LALM student trained using the SightSound-R1 framework is capable of achieving reasoning capabilities from the LVLM teacher, supervised by a GPT-4o-audio Fact-Checker.}
    \label{fig:pipeline}
    \vspace{-0.3cm}
\end{figure*}

\begin{itemize}[noitemsep,topsep=0pt]
\item We identify and analyze the performance gap between LVLMs and LALMs in multi-step reasoning on AVQA benchmark.
\item We propose \textbf{SightSound-R1}, a structured and automatic pipeline that elicits, verifies, and distills audio-focused CoT from LVLMs into an LALM through SFT and GRPO.
\item We demonstrate strong and generalizable results: SightSound-R1 improves LALM reasoning on AVQA and achieves competitive accuracy on unseen MMAU and MUSIC-AVQA benchmarks.
\end{itemize}



\vspace{-0.2cm}

\section{Related Work}
\label{sec:format}

\subsection{Language and Multimodal Reasoning}
\label{ssec:subhead}
Recent foundation models, including the Qwen series \cite{qwen3,qwen25vl,qwen25omni,qwen2audio}, ChatGPT \cite{Gpt-4o,o1}, and Gemini \cite{gemini}, have demonstrated remarkable general reasoning capabilities across diverse domains. The DeepSeek series \cite{deepseek,deepseekmath} in particular has drawn attention for improving reasoning through reinforcement learning with verifiable rewards (RLVR), an approach successfully extended to vision, audio, and audio–visual settings \cite{vision-r1,Visual-rft,r1-aqa,Omni-R1,echoink}.

In the audio domain, recent work has introduced explicit chain-of-thought (CoT) supervision for large audio–language models (LALMs). Audio-CoT \cite{Audio-cot} extends chain-of-thought reasoning to LALMs, while Audio-Reasoner \cite{Audio-reasoner} introduces structured multi-stage reasoning processes for audio understanding. AudSemThinker \cite{audsemthinker} further contributes to this line of work by developing semantic reasoning frameworks. Other efforts, such as SARI \cite{sari} and Audio-Thinker \cite{Audio-Thinker}, pair audio captions with reasoning LLMs to build corpora for supervised fine-tuning (SFT) and GRPO training. Beyond audio-only reasoning, cross-modal audio–visual approaches are emerging: ThinkSound \cite{thinksound} integrates CoT reasoning from both LVLMs and LALMs for video-to-audio generation, while Ren et al. \cite{hearing} elicit audio captions from silent video on LVLM to support text-assisted video-to-audio synthesis. Jiang et al. \cite{bridging} further highlight the promise of distilling LVLM outputs into LALMs for visible-sound recognition tasks.
\subsection{Reasoning Distillation}
\label{ssec:subhead}
Reasoning distillation represents a key direction for making advanced reasoning capabilities accessible to smaller models.
NaturalThoughts \cite{naturalthoughts} curates systematic reasoning traces from strong teachers, while Aurelia \cite{aurelia} proposes training-free, test-time reasoning distillation for audio–visual LLMs using multi-agent frameworks. MiCoTA \cite{micota} and Li et al. \cite{small} tackle the learnability gap in small language models (SLMs), introducing strategies that balance reasoning complexity to enable effective distillation.  Nonetheless, all these methods distill reasoning from stronger to weaker models of \textit{the same modality}. In the audio domain, unfortunately, strong audio reasoners are still in demand, not to mention the absence of large audio reasoning datasets. Therefore, this work is proposed to circumvent the lack of such strong audio teachers and datasets by the ``free lunch'' of vision teachers on audio-visual datasets.

\section{Method}
\label{sec:method}

SightSound-R1 distills stepwise reasoning from a strong LVLM into a student LALM via three stages (Fig.~\ref{fig:pipeline}). 
(1) \emph{Teacher reasoning generation:} we elicit multiple audio-focused CoT traces from silent video using test-time scaling with self-consistency. 
(2) \emph{Audio-grounded fact verification (AGFV):} a lightweight checker validates the teacher’s audio claims against the true audio and filters hallucinated traces to curate a fact-checked corpus. 
(3) \emph{Student training:} the LALM first undergoes SFT on verified CoT to learn format and alignment, then GRPO to optimize answer accuracy and CoT format with a KL-regularized objective. 
This pipeline transfers vision reasoning traces while preserving audio grounding, and it scales to arbitrary audio–visual scenes \textit{without} human CoT annotations!

\subsection{Teacher Reasoning Generation with Test-Time Scaling}
\label{ssec:teacher}
Test-time scaling with self-consistency elicits multiple diverse CoT traces and keeps only consensus/unanimous answers, reducing hallucinations while improving supervision quality without extra samples. 
We use Qwen2.5-VL-32B-Instruct as the LVLM teacher $\mathcal{T}$ to produce multiple \emph{audio-focused} CoT traces from \emph{silent} video. Given a video--question pair $(v,q)$ and an audio-focused prompt $P_{\text{audio}}$, we sample $n$ traces via self-consistency:
\[
\begin{aligned}
\mathcal{R}(v,q) &= \{\,r_i=\mathcal{T}(P_{\text{audio}}(v,q))\,\}_{i=1}^{n}, \\
\mathcal{A} &= \{\text{ExtractAnswer}(r_i)\}_{i=1}^{n}.
\end{aligned}
\]
We retain a sample when teacher answers are unanimous:
\[
|\text{unique}(\mathcal{A})|=1 \;\Rightarrow\; (v,q,\mathcal{R})\in\mathcal{D}_{\text{reason}}.
\]
This generation strategy (Fig.~\ref{fig:pipeline}) hence yields diverse but high-confidence reasoning trajectories.


\subsection{Audio-Grounded Fact Verification}
\label{ssec:factcheck}

Because the LVLM teacher cannot hear, its CoT traces may hallucinate with non-existent sounds. We verify each trace $r\in\mathcal{R}$ against the true audio $a$ using an audio checker $\mathcal{C}$ (another LALM e.g., GPT-4o-audio) with a binary decision:
\[
\text{AGFV}(r,a)=
\begin{cases}
\text{accept}, & \text{if }\mathcal{C}(r,a)=\text{``yes''} \\
\text{reject}, & \text{otherwise.}
\end{cases}
\]
Accepted traces form a fact-checked corpus
\(
\mathcal{D}_{\text{FC}}=\{(a,q,r): r\in\mathcal{R},\ \text{AGFV}(r,a)=\text{accept}\}.
\)
This step filters hallucinated claims while remaining model-agnostic. For cost-sensitive settings, $\mathcal{C}$ can be replaced by an audio–text verifier trained for binary alignment using a small calibration set, similar to the binary switch in \cite{bridging}.




\subsection{Student Training on Distilled Teacher Output}
\label{ssec:student}

We train the LALM student Qwen2-Audio-7B-Instruct in two stages: SFT on verified CoT followed by GRPO with a KL anchor. \textbf{SFT} bootstraps the student with teacher CoT format, while \textbf{GRPO} further optimizes CoT format and answer accuracy with teacher label via exploration, achieving robust, generalizable reasoning.

Let $\mathbf{x}=(a,q)$ denote the audio–question pair and $\mathbf{y}=r$ be a fact-checked trace from $\mathcal{D}_{\text{FC}}$. We optimize only LoRA \cite{hu2022lora} parameters $\theta_{\text{LoRA}}$ on top of a frozen base $\theta_{\text{base}}$. The SFT objective is:

\begin{align}
\mathcal{L}_{\text{SFT}}(\theta_{\text{LoRA}}) &= \mathbb{E}_{(\mathbf{x},\mathbf{y})\sim \mathcal{D}_{\text{FC}}} \left[-\sum_{t=1}^{|\mathbf{y}|}\log \pi_{\theta_{\text{base}}\oplus\theta_{\text{LoRA}}}(y_t\,|\,\mathbf{x},\mathbf{y}_{<t})\right] \nonumber \\
\end{align}

Following DeepSeekMath \cite{deepseekmath}, we then sample $G$ responses $\{o_1, o_2, \ldots, o_G\}$ from the current policy $\pi_{\theta_{\text{old}}}$ and compute their rewards $\{r_1, r_2, \ldots, r_G\}$. The advantage is normalized as 
$\hat{A}_{i,t} = \tilde{r}_i = \dfrac{r_i - \text{mean}(r)}{\text{std}(r)}$, 
where $r_i \in \{0, 1, 2\}$ is the reward for response $o_i$. The policy model $\pi_\theta$ is optimized via the clipped objective:
\begin{equation}
\begin{split}
J_{\text{GRPO}}(\theta) = \mathbb{E}_{q \sim P(Q), \{o_i\}_{i=1}^G \sim \pi_{\theta_{\text{old}}}(O|q)} \Bigg[ \frac{1}{G} \sum_{i=1}^{G} \frac{1}{|o_i|} \sum_{t=1}^{|o_i|} \\
\Bigg( \min \left[ \frac{\pi_\theta(o_{i,t}|q, o_{i,<t})}{\pi_{\theta_{\text{old}}}(o_{i,t}|q, o_{i,<t})} \hat{A}_{i,t}, \right. \\
\left. \text{clip}\left(\frac{\pi_\theta(o_{i,t}|q, o_{i,<t})}{\pi_{\theta_{\text{old}}}(o_{i,t}|q, o_{i,<t})}, 1-\epsilon, 1+\epsilon\right) \hat{A}_{i,t} \right] \\
- \beta D_{\text{KL}}[\pi_\theta || \pi_{\text{ref}}] \Bigg) \Bigg]
\end{split}
\end{equation}
where $o_{i,t}$ denotes the $t$-th token in response $o_i$, $\epsilon$ is the clipping parameter, $\beta$ controls KL divergence strength from the reference policy $\pi_{\text{ref}}$, and $\pi_{\theta_{\text{old}}}$ is the policy before the current update. The reward function $r_i$ evaluates format compliance and answer correctness:

\begin{itemize}[noitemsep,topsep=0pt]
\item \textbf{Accuracy reward} (+1): if the answer matches the teacher’s prediction.
\item \textbf{Format reward} (+1): if the output correctly uses 
\texttt{<think>} \dots \texttt{</think>} and 
\texttt{<answer>} \dots \texttt{</answer>} tags.
\item Otherwise, the reward is 0.
\end{itemize}


\label{sec:typestyle}

\begin{table}[t]
\centering
\rowcolors{2}{lightgray}{white}
\begin{adjustbox}{max width=\columnwidth}
\begin{tabular}{ccc}
\hline
\textbf{Models} & \textbf{Method} & \textbf{Accuracy} \\
\hline
\multicolumn{3}{l}{\textit{Large Audio–Language Models}}
\\
\hline
Qwen2.5-Omni-3B (audio) & Direct Inference & 73.6 \\
Qwen2.5-Omni-7B (audio) & Direct Inference & 74.7 \\
Qwen2-Audio-7B-Instruct & Direct Inference & 67.1 \\
Qwen2-Audio-7B-Instruct & Zero-Shot-CoT & 57.7 \\
Qwen2-Audio-7B-Instruct & Test-time CoT-distill & 82.7 \\
Qwen2-Audio-7B-Instruct & SFT (w/. Ground Truth) & \textbf{86.5} \\
\hline
\rowcolor{white}\multicolumn{3}{l}{\textit{Large Vision–Language Models}} \\
\hline
Qwen2.5-Omni-3B (video) & Direct Inference & 86.5 \\
Qwen2.5-Omni-7B (video) & Direct Inference & \textbf{87.4} \\
Qwen2.5-VL-7B-Instruct & Direct Inference & 85.7 \\
Qwen2.5-VL-32B-Instruct & Direct Inference & 85.8 \\
Qwen2.5-VL-32B-Instruct & Zero-Shot-CoT & 84.6 \\
Qwen2.5-VL-32B-Instruct & Zero-Shot-Audio-CoT & 85.5 \\
\hline
\multicolumn{3}{l}{\textit{Large Audio\&Visual-Language Models}} \\
\hline
Qwen2.5-Omni-3B & Direct Inference & \textbf{88.5} \\
Qwen2.5-Omni-7B & Direct Inference & \textbf{89.5} \\
\hline
\end{tabular}
\end{adjustbox}
\caption{Accuracies (\%) on the AVQA validation dataset. Scores are computed by exact option letter match; when absent, textual responses are aligned to the closest option via content similarity.}
\label{table:preliminary}
\vspace{-0.2cm}
\end{table}

As shown in Figure~\ref{fig:pipeline}, this pipeline allows the student to internalize the teacher’s reasoning patterns while continuing to explore with its own audio perceptors. The approach scales naturally across diverse audio–visual settings analyzed and annotated by strong, test-time-scaled LVLM teacher.

\begin{table*}[t]
\centering
\rowcolors{2}{lightgray}{white}
\begin{adjustbox}{max width=\linewidth}
\begin{tabular}{cccccc|cccccc}
\hline
\textbf{Models} & \textbf{Method} & \multicolumn{4}{c}{\textbf{MMAU Test-mini}} & \multicolumn{5}{c}{\textbf{MUSIC-AVQA Test}} \\
\hline
& & \textbf{Sound} & \textbf{Speech} & \textbf{Music} &  \textbf{Avg.} & \textbf{Temp} & \textbf{Comp} & \textbf{Cnt} & \textbf{Exist}& \textbf{Avg.}\\
\hline
\rowcolor{white}\multicolumn{11}{l}{\textit{\textbf{Baselines:}}} \\
Qwen2-Audio-7B-Instruct & Direct Inference & 64.3 & 52.6 & 61.7 & 59.5 & 57.2 & 57.8 & 55.7 & 55.1 & 55.6\\
Qwen2-Audio-7B-Instruct & SFT (Ground Truth Label) & 66.7 & 50.8 & 61.1 & 59.5 & 60.4 & 62.7 & \textbf{61.1} & \textbf{60.7} & \textbf{61.1}\\
Qwen2-Audio-7B-Instruct & SFT (Distilled Label)  & 64.3 & 52.3 & 60.5 & 59.0 & 59.5 & 61.4 & 59.1 & 60.2 & 58.8\\
Qwen2-Audio-7B-Instruct & GRPO (Distilled Label)  & 62.5 & 49.8 & 59.3 & 57.2 & 59.3 & 60.8 & 59.1 & 60.3 & 58.8\\
Audio-Thinker \cite{Audio-Thinker} & SFT (Ground Truth CoT) 
  & 63.4 & 56.3 & 54.4 & 57.8 & -- & -- & -- & -- & -- \\
\textemdash & GRPO (Ground Truth CoT) 
  & \textbf{70.3} & \textbf{61.6} & \textbf{63.2} & \textbf{65.0} & -- & -- & -- & -- & -- \\

\hline
\multicolumn{3}{l}{\textit{\textbf{Ours (SFT/GRPO on Distilled CoT):}}}
\\
Qwen2-Audio-7B-Instruct & SFT & 61.3 & 47.1 & 48.5 & 52.3 & 56.7 & 57.7 & 56.5 & 54.3 & 55.1\\
Qwen2-Audio-7B-Instruct & AGFV + SFT & 63.1 & 47.7 & 51.5 & 54.1 & 58.2 & 59.4 & 56.9 & 56.5 & 56.5\\
Qwen2-Audio-7B-Instruct & TTS + AGFV + SFT & 61.6 & 48.3 & 50.6 & 53.5 & 60.6 & 59.0 & 59.1 & 57.5 & 58.2\\
\textbf{SightSound-R1} & TTS + AGFV + SFT + GRPO & \textbf{\underline{66.1}} & 49.8 & 52.7 & 56.2 & \textbf{\underline{62.7}} & \textbf{\underline{63.3}} & 60.1 & 59.7 & \textbf{\underline{59.5}}\\
\hline
\end{tabular}
\end{adjustbox}
\caption{Accuracies (\%) on the MMAU Test-mini (v05.15.25) and MUSIC-AVQA Testset. MUSIC-AVQA sub-categories: \textbf{Temp} = Temporal, \textbf{Comp} = Comparative, \textbf{Cnt} = Counting, \textbf{Exist} = Existential. Abbreviations: \textbf{SFT} = Supervised Fine-tuning, \textbf{GRPO} = Group Relative Policy Optimization, \textbf{TTS} = Test-time Scaling, 
\textbf{AGFV} = Audio-Grounded Fact Verification.}
\label{table:comparsion}
\vspace{-0.2cm}
\end{table*}

\section{Experiments}

\subsection{Dataset and Implementation Details}
We evaluate on AVQA, MMAU, and MUSIC-AVQA. AVQA \cite{avqa} provides large-scale audio–visual QA; following R1-AQA \cite{r1-aqa}, we create an audio–text variant by extracting the audio track and replacing “video’’ with “audio’’ in questions, while retaining paired silent-video prompts for LVLM reasoning. For MMAU \cite{mmau}, we report on the official \textit{test-mini} (v05.15.25) with 1k audio QA pairs. MUSIC-AVQA \cite{musicavqa} contains 7k QA from YouTube performance videos spanning 22 instruments and 9 question types. We report per-category (Table~\ref{table:comparsion}) and overall accuracy.

All experiments used the SWIFT framework \cite{swift} on a single node with 8 NVIDIA A40 GPUs, applying LoRA \cite{hu2022lora} for parameter-efficient fine-tuning and vLLM \cite{vllm} for accelerated rollout generation during GRPO. In the SFT stage, Qwen2-Audio-7B-Instruct was trained with per-GPU batch size 8, learning rate $5\times10^{-5}$, LoRA rank 8 with $\alpha=16$, for 2000 steps. During GRPO, we switched to full-parameter tuning, allocating 2 GPUs for rollout generation and 6 GPUs for policy optimization, sampling 8 completions per input prompt (192 candidate responses per step) with per-device batch size 4, learning rate $1\times10^{-6}$, temperature 1.0, and KL coefficient $\beta=0.04$ for up to 1000 steps. Best checkpoints are selected by validation accuracy for both stages. \textit{All training is performed on the AVQA dataset.}

\subsection{Results and Analysis}
\textbf{Preliminary:} We first justified the existence of the performance gap between LALMs and LVLMs. On the AVQA validation set (Table~\ref{table:preliminary}), LVLMs substantially outperform LALMs, with multimodal Qwen2.5-Omni (3B/7B) leading at 88–89\% while Qwen2-Audio-7B trails under direct inference. Zero-Shot-CoT further degrades Qwen2-Audio-7B (57.7\% vs.\ 67.1\%), indicating its backbone lacks reasoning capacity for helpful CoT. By contrast, Qwen2.5-VL-32B maintains strong performance generating audio-focused CoT from silent video (85.5\%; see Figure~\ref{fig:pipeline}). Using Qwen2.5-VL-32B's Audio-CoT for test-time distillation raises Qwen2-Audio-7B to 82.7\%, while supervised training with ground-truth labels reaches 86.5\%, supporting our hypothesis that LVLM's reasoning can be transferred to audio models. These findings ground \textbf{SightSound-R1} to distill stronger LVLM reasoning into a weaker LALM.

On MMAU Test-mini (Table~\ref{table:comparsion}), SightSound-R1 achieves strongest performance on Sound tasks (66.1\%), followed by Music (52.7\%) and Speech (49.8\%). Although this underperforms the Audio-Thinker \cite{Audio-Thinker} baseline on Sound tasks (66.1\% vs.\ 70.3\%), our method relies solely on LVLM teacher \textit{without ground-truth signals}, effectively demonstrating scalable cross-modal knowledge transfer.

On MUSIC-AVQA test set, SightSound-R1 achieves 59.5\% accuracy and ranks second-best overall, surpassing baselines in Temporal (62.7\%) and Comparative (63.3\%) reasoning tasks. Our ablation studies reveal progressive component contributions: base SFT on distilled CoT alone yields 52.3\% on MMAU, adding Audio-Grounded Fact Verification (AGFV) improves performance to 54.1\% by filtering hallucinations, and the full SightSound-R1 with GRPO achieves significant improvements on both MMAU sound tasks and MUSIC-AVQA. Test-Time Scaling (TTS) shows mixed effects, slightly decreasing MMAU performance but improving MUSIC-AVQA results, demonstrating robust transfer of reasoning patterns.

Notably, SFT on distilled labels maintains zero-shot performance on MMAU and improves results on MUSIC-AVQA, indicating that LVLM label distillation yields a stable performance distribution in audio–visual scenes. In contrast, our distilled-CoT approach trades accuracy on Speech and Music for interpretable reasoning, stronger sound-event perception, and scalability. GRPO on distilled labels performs worse than SFT because the base model is already tuned to output only labels and cannot explore diverse reasoning traces in the \texttt{<think>\ldots</think>} format.

Moreover, the performance pattern of SightSound-R1 (Sound improved, Speech and Music decreased) highlights both the strengths and limits of cross-modal reasoning transfer. LVLMs readily infer visible sound events but fail to capture fine acoustic properties such as tempo, pitch, timbre, or spoken content, which lack clear visual correlates. As a result, SFT on these CoT traces can misguide the student toward deviations or hallucinations. Future work should better integrate with LALM perception to achieve robust reasoning.

\section{Conclusion}
\label{sec:typestyle}
We introduced \textbf{SightSound-R1}, a cross-modal distillation framework that transfers LVLM reasoning to LALM by AVQA. By combining test-time scaling, audio-grounded fact verification, and SFT/GRPO on distilled CoT, SightSound-R1 achieves \(66.1\%\) on MMAU Test-mini Sound and \(59.5\%\) on MUSIC-AVQA, showing stronger gains on unseen audio–visual reasoning tasks than label distillation. Our framework is naturally scalable on audio-visual data without human CoT annotation. In the future, we plan to extend the framework to more abundant Internet videos (e.g., YouTube).
\vspace{-0.3cm}

\section{Acknowledgement}
This work is funded by the National Institutes of Health (NIH-NIDCD) and a grant from Marie-Josee and Henry R. Kravis.

\vfill\pagebreak

\bibliographystyle{IEEEbib}
\bibliography{strings,refs}

\end{document}